\documentstyle[11pt,epsf]{article}

\newcommand{\beq}{\begin{equation}}
\newcommand{\eeq}{\end{equation}}
\newcommand{\beqa}{\begin{eqnarray}}
\newcommand{\eeqa}{\end{eqnarray}}

\newcommand{\NPB}[1]{{\it Nucl. Phys.}\ {\bf B{#1}}}
\newcommand{\PLB}[1]{{\it Phys. Lett.}\ {\bf B{#1}}}

\newcommand{\PRL}[1]{{\it Phys. Rev. Lett.}\ {\bf #1}}

\newcommand{\lae}{\begin{array}{c} < \vspace{-.85em} \\ {\scriptstyle \sim}
\end{array}}
\newcommand{\gae}{\begin{array}{c} > \vspace{-.85em} \\ {\scriptstyle \sim}
\end{array}}
\newcommand{\laef}{\begin{array}{c} \vspace{.5em}{\scriptstyle <} 
\vspace{-1.3em} \\ \hspace{.5mm}{\scriptscriptstyle \sim} \end{array}}

\begin{document}

\begin{titlepage}
\def\thepage {}        

\title{\bf Flavor Physics and the Triviality Bound \\ 
on the Higgs Mass}

\author{
R.~S.~Chivukula\thanks{e-mail addresses: sekhar@bu.edu, dobrescu@budoe.bu.edu,
simmons@bu.edu}, B.~A.~Dobrescu$^*$, and E.~H.~Simmons$^*$ 
\vspace{3mm}\\
{\it Department of Physics, Boston University,} \\
{\it 590 Commonwealth Ave., Boston MA  02215}}

\date{February 24, 1997}

\maketitle

\bigskip
\begin{picture}(0,0)(0,0)
\put(295,250){BUHEP-97-6}
\put(295,235){hep-ph/9702416}
\end{picture}
\vspace{24pt}

\begin{abstract}

  The triviality of the scalar sector of the standard one-doublet Higgs
  model implies that this model is only an effective low-energy theory
  valid below some cut-off scale $\Lambda$. The underlying high-energy
  theory must include flavor dynamics at a scale of order $\Lambda$ or
  greater in order to give rise to the different Yukawa couplings of the
  Higgs to ordinary fermions.  This flavor dynamics will generically
  produce flavor-changing neutral currents and non-universal corrections
  to $Z\to b\bar{b}$. We show that the experimental
  constraints on the neutral $D$-meson mass
  difference imply that $\Lambda$ must be greater than of order 21 TeV.
  We also discuss bounds on $\Lambda$ from the 
  constraints on extra contributions to the $K_L$-$K_S$ mass difference
  and to the coupling of the $Z$ boson to $b$-quarks.
  For theories defined about the
  infrared-stable Gaussian fixed-point, we estimate that this lower
  bound on $\Lambda$ yields an upper bound of approximately 460 GeV on
  the Higgs boson's mass, independent of the regulator chosen to define
  the theory.

\pagestyle{empty}
\end{abstract}
\end{titlepage}


\section{Introduction}

The triviality \cite{triviality} of the scalar sector of the standard
one-doublet Higgs model implies that this theory is only an effective
low-energy theory valid below some cut-off scale $\Lambda$.  Physically
this scale marks the appearance of new strongly-interacting
symmetry-breaking dynamics, examples of which include ``top-mode''
standard models \cite{topmode} and composite Higgs models \cite{chiggs}.
As the Higgs mass, $M_H$, increases, the upper bound on the scale of new
physics decreases.  Thus, if one requires that $M_H / \Lambda$ be small
enough to afford the effective Higgs theory some range of validity (or
to minimize the effects of regularization in the context of a
calculation in the scalar theory), one arrives at the conventional upper
limit on $M_H$ of approximately 700 GeV \cite{refbound}.

In a previous paper \cite{higgsmass}, two of us discussed how
constraints on custodial symmetry violation affect the upper bound on
the Higgs mass.  We noted that the underlying high-energy physics must
provide some custodial symmetry violation in order to explain the
large mass splitting between the top and bottom quarks.  This enabled us
to show that the experimental constraint
on the amount of custodial symmetry violation, $\vert\Delta \rho_* \vert
= \alpha \vert T \vert$, implies that the scale $\Lambda$ must be
greater than of order 7.5 TeV, and we argued that the bound is
regularization-independent.\footnote{The $S$ parameter \cite{Spar} also
  provides a limit on $\Lambda$, but it is weaker than that from $T$.
  The only dimension-6 operator that contributes \cite{wyler}
  to $S$ is $\frac{1}{\Lambda^2} \left\{ \left[ D_\mu,
  D_\nu \right] \phi \right\}^\dagger \left[ D_\mu, D_\nu \right] \phi
  ~.$ Since this implies $S = 2 \pi v^2 / \Lambda^2$, the 95 \% c.l. $S
  \lae 0.23$ \cite{data} yields $\Lambda \gae 1.3 \, {\rm TeV}$.}

This lower bound on the scale $\Lambda$ yielded \cite{higgsmass} an
upper limit of approximately 550 GeV on the Higgs boson's mass.

Similarly, regardless of the precise nature of the underlying
strongly-interacting physics, there must be flavor dynamics at a scale
of order $\Lambda$ or greater that gives rise to the different Yukawa
couplings of the Higgs boson to ordinary fermions.  As in extended
technicolor theories \cite{Lane,CSS}, this flavor dynamics will
generically cause flavor-changing neutral currents and non-universal
corrections to the decay $Z\to b\bar{b}$. In this note we derive a lower
bound on $\Lambda$ from the experimental constraints on extra
contributions to the neutral meson mass differences and to the coupling
of the $Z$ boson to $b$-quarks.  We then estimate the upper limit on the
Higgs boson's mass corresponding to this lower bound on $\Lambda$.

Since the operators responsible for generating quark masses and for
causing flavor-changing neutral currents violate flavor symmetries
differently \cite{ctsm}, in principle one could construct a theory with
an approximate GIM symmetry \cite{ctsm,oldtcgim,technigim}.  In such models,
flavor-changing neutral currents would be suppressed but different
quarks would still receive different masses.  A theory of this type
which included a light scalar state (unlike the examples
\cite{ctsm,oldtcgim,technigim}) would be able to evade the flavor-changing neutral current
limits discussed here.  However, such models would still \cite{CSS} be
subject to the bounds we find from $Z \to b\bar{b}$ and could also give
rise to potentially dangerous contributions to other processes
\cite{rc}.

Implicitly assumed in these bounds is the naive scaling that one expects
near the infrared-stable Gaussian fixed point of scalar field theory.
Other fixed points with very different scaling behavior may also exist.
In this case, the bounds we discuss here would not apply.  However, as
discussed in \cite{higgsmass}, to construct a phenomenologically viable
theory of a strongly-interacting Higgs sector it is not sufficient to
simply construct a theory with a heavy Higgs boson. To be consistent
with the experimental bound on $\vert\Delta\rho_*\vert$, one must also
ensure that all potentially custodial-isospin-violating operators remain
irrelevant.  For this reason, we expect constructing a
phenomenologically-acceptable non-trivial scalar electroweak symmetry
breaking sector to be difficult.  To our knowledge, no acceptable
model of this sort has been proposed.

\section{Flavor Physics and the Higgs Couplings}
\setcounter{equation}{0}

In what follows, we consider a theory with an arbitrary
strongly-interacting sector which reduces at low energies to the
one-Higgs-doublet standard model.  Our goal is to understand how the
underlying strongly-interacting dynamics would manifest itself in
low-energy flavor physics.  

To estimate the sizes of various effects of the underlying physics, we
rely on dimensional analysis. As noted by Georgi \cite{generalized}, a
theory\footnote{These dimensional estimates only apply if the low-energy
  theory, when viewed as a scalar field theory, is defined about the
  infrared-stable Gaussian fixed-point. As discussed above, we expect
  this to be the case.} with light scalar particles belonging to a
single symmetry-group representation depends on two parameters:
$\Lambda$, the scale of the underlying physics, and $f$ (the analog of
$f_\pi$ in QCD), which measures the amplitude for producing the scalar
particles from the vacuum. Our estimates of the sizes of the low-energy
effects of the underlying physics will depend on the ratio $\kappa
\equiv \Lambda / f$, which determines the sizes of coupling constants in
the low-energy theory. The value of $\kappa$ is expected to fall between
1 and $4\pi$.  For example, in QCD we find that the $\rho$-coupling is
$g_\rho={\cal O}(\kappa)\approx 6$.  In a QCD-like theory with $N_c$
colors and $N_f$ flavors one expects \cite{reconsider} that
\beq \kappa \approx \min \left({4\pi a\over N_c^{1/2}}, {4\pi b\over
  N_f^{1/2}}\right)~, 
\eeq 
where $a$ and $b$ are constants of order 1.
In the results that follow, we will display the dependence on $\kappa$
explicitly; when giving numerical examples, we set $\kappa$ equal to the
geometric mean of 1 and $4\pi$, {\it i.e.} $\kappa \approx 3.5$.

We begin by considering what the observed masses of the ordinary
fermions imply about the underlying physics. Providing the different
masses of the fermions requires flavor physics (analogous to
extended-technicolor interactions (ETC) \cite{Lane,etc}) which couples
the left-handed quark doublets $\psi_L$ and right-handed singlets $q_R$
to the strongly-interacting ``preon'' constituents of the Higgs doublet.
At low energies, these interactions produce the quark Yukawa couplings.
Assuming, for simplicity, that these new flavor interactions are gauge
interactions with gauge coupling $g$ and gauge boson mass $M$,
dimensional analysis \cite{dima} allows us to estimate that the size of
the resulting Yukawa coupling is \cite{chiggs} of order
$(g^2/M^2)(\Lambda^2/\kappa)$, i.e.
\beq {\lower35pt\hbox{\epsfysize=1.0 truein
    \epsfbox{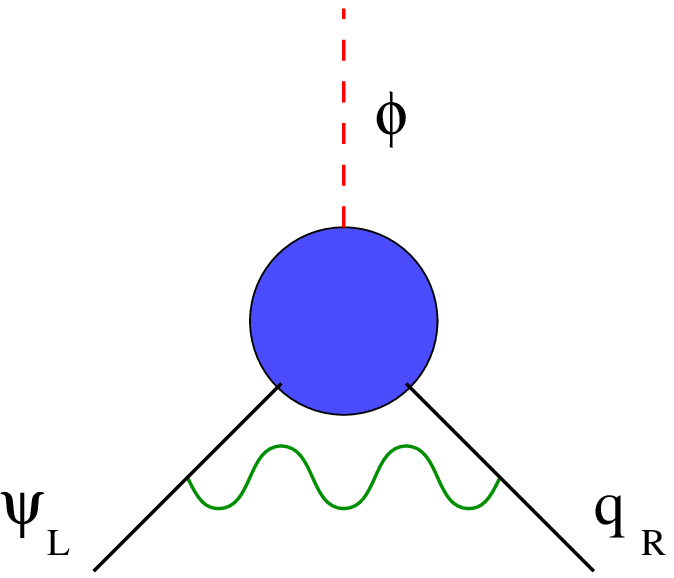}}} \Rightarrow {g^2 \over M^2} {\Lambda^2 \over
  \kappa}\bar{q}_R \phi \psi_L ~.
\label{eq:quarkpreon}
\eeq
In order to give rise to a quark mass $m_q$, the 
Yukawa coupling must be equal to
\beq
{\sqrt{2} m_q \over v}
\eeq
where $v\approx 246$ GeV. This implies\footnote{Because the low-energy
  theory  is (approximately) the standard
model, unitarity in the scattering amplitude $q\bar{q}\to
W_L W_L$ is ensured due to Higgs Boson exchange.
In this case, unlike ref. \cite{appchan}, there is no
{\it upper bound} on the scale $M/g$.}
\beq
\Lambda \gae {M \over g} \sqrt{\sqrt{2} \kappa {m_q \over v}}~.
\label{eq:yukawa}
\eeq 
Thus, if we set a lower limit on $M/g$ from low-energy flavor physics,
eqn.(\ref{eq:yukawa}) will give a lower bound on $\Lambda$.

The high-energy flavor physics responsible for the generation of the
quark-preon couplings {\it must} distinguish between different flavors
so as to give rise to the different masses of the corresponding
fermions.  In addition to the Higgs-fermion coupling discussed above,
the flavor physics will also give rise to flavor-specific couplings of
ordinary fermions to themselves \cite{Lane} and of weak currents of
ordinary fermions to weak currents of preons \cite{CSS}.  Such
interactions will cause potentially visible effects on flavor physics at
low energies.  For example, the interaction between weak currents of
preons and ordinary fermions gives rise to an operator that can alter
the $Zb\bar b$ vertex.  If the new gauge interactions commute with
$SU(2)_W$, {\it i.e.} if the gauge bosons do not carry weak charge,
using dimensional analysis we find the coefficient of the appropriate
operator in the effective Lagrangian to be
\beq
{\lower35pt\hbox{\epsfysize=1.0 truein \epsfbox{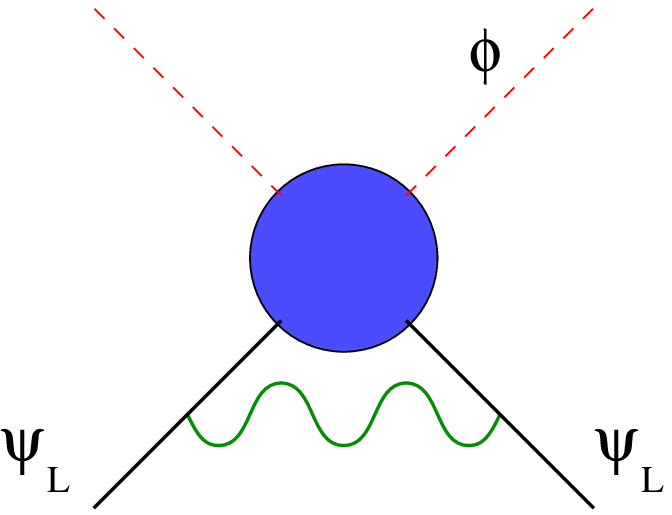}}} 
\Rightarrow 
-2 \frac{g^2}{M^2} \left[ (D_\mu \phi)^\dagger i \frac{\tau_3}{2} \phi -
\phi^\dagger i \frac{\tau_3}{2} D_\mu \phi \right] \overline q_{L}
\gamma^{\mu} \frac{\tau_3}{2} q_{L}~.
\label{eq:zbb}
\eeq
After spontaneous symmetry breaking, this shifts the coupling of the $Z$ to $q\bar{q}$
\beq
\delta g_L^q = - \frac{g^2}{M^2} \frac{v^2}{2}T^q_3~,
\label{delg}
\eeq 
where $T^q_3 = \pm \frac{1}{2}$ is the weak-isospin of the quark.  Note
that, as in the case of conventional ``commuting'' ETC
\protect\cite{CSS} models, the sign of the coupling shift is prescribed
and the shift tends to reduce the decay-width of $Z$ to each quark.

The new current-current interactions among ordinary fermions, on the
other hand generically give rise to flavor-changing neutral currents (as
previously noted in \cite{Lane} for the case of ETC theories) that
affect Kaon, $D$-meson, and $B$-meson physics.  
For instance, consider the interactions responsible
for the $s$-quark mass. Through Cabibbo mixing, these interactions must
couple to the $d$-quark as well.  This will generally give rise to the
interactions
\beqa
{\cal L}_{eff} & = & - \, (\cos \theta_L^s \sin \theta_L^s)^2 
\frac{g^{2}}{M^{2}}
( \overline s_L  \gamma^{\mu} d_L )(\overline s_L  \gamma_{\mu} d_L)
\nonumber \\ [2mm]
& & -  \, (\cos \theta_R^s \sin \theta_R^s)^2 
\frac{g^{2}}{M^{2}}
( \overline s_R \gamma^{\mu} d_R)(\overline s_R \gamma_{\mu} d_R)
\nonumber \\ [2mm]
& & - \,
\cos \theta_L^s \sin \theta_L^s \cos \theta_R^s \sin \theta_R^s
\frac{g^{2}}{M^{2}}
( \overline s_L  \gamma^{\mu} d_L )(\overline s_R \gamma_{\mu} d_R)~,
\label{ops1}
\eeqa
where the coupling $g$ and mass $M$ are of the same order as those
in the interactions which ultimately give rise to the $s$-quark
Yukawa coupling in eqn.~(\ref{eq:quarkpreon}), and the angles
$\theta^s_L$ and $\theta^s_R$ represent the relation between the
gauge eigenstates and the mass eigenstates.  The operators in
eqn.~(\ref{ops1}) will clearly affect neutral Kaon physics.  Similarly, the
interactions responsible for other quarks' masses will give rise to
operators that contribute to mixing and decays of various mesons.

\section{Constraints on $\Lambda$ from Flavor Physics}
\setcounter{equation}{0}

\subsection{Flavor-Changing Neutral Currents: $\Delta S$ and $\Delta B$}

To start, let us consider the
four-fermion interactions in eqn.~(\ref{ops1}), which will alter the
predicted value of the $K_L - K_S$ mass difference.  Using the
vacuum-insertion approximation \cite{vacinsert}, we can estimate
separately how much the purely left-handed (LL), purely right-handed
(RR) and mixed (LR) current-current operators contribute.  Requiring
each contribution to be less than the observed mass difference $\Delta
m_K$, we find the bounds
\beqa
\left(\frac{M}{g}\right)_{\! {\rm LL,RR}} \! & \gae &\!
f_K \left( \frac{2  m_K B_K}{3 \Delta m_K }\right)^{\! 1/2}
\cos \theta_{L,R}^s \sin \theta_{L,R}^s
\\ [2mm]
& \approx & \! 0.92 \times 10^{3} \, {\rm TeV} 
\cos \theta_{L,R}^s \sin \theta_{L,R}^s
\eeqa
from the first two operators in eqn.~(\ref{ops1}), and
\beqa
\hspace*{-8mm}
\!\!\left(\frac{M}{g}\right)_{\! {\rm LR}}\!\!
&\!\gae\! & \!
f_K \left\{ \frac{m_K B_K^{\prime}}{3 \Delta m_K } 
\left[ \frac{m_K^2}{(m_s + m_d)^2} - \frac{3}{2} \right] 
\right\}^{\! 1/2}\!\!\!
(\cos \theta_L^s \sin \theta_L^s \cos \theta_R^s \sin \theta_R^s)^{\! 1/2}
\\ [2mm]
& \! \approx \! & \! 1.4 \times 10^{3} \, {\rm TeV} \,
(\cos \theta_L^s \sin \theta_L^s \cos \theta_R^s \sin \theta_R^s)^{\! 1/2}
\eeqa
from the last operator in eqn.~(\ref{ops1}). In evaluating these
expressions, we have used $f_K \approx 113$ MeV, the ``bag''
factors $B_K, B_K^\prime \sim 0.7$, and $m_s + m_d \sim 200$ MeV.  In
order to produce the observed $d - s$ mixing, we expect that at least one of the
angles $\theta_L^s,\ \theta_R^s$ is of order the Cabibbo angle,
$\theta_C$.  Then we find from any one operator 
\beq
\frac{M}{g}
\gae 200 \, {\rm TeV}~.
\label{fcncmbound}
\eeq
From eqn.~(\ref{eq:yukawa}) it follows that
\beq
\Lambda \gae 6.8 \, {\rm TeV} 
\sqrt{\kappa\left({m_s\over 200\, {\rm MeV}}\right)}~.
\label{eq:fcncbound}
\eeq
For $\kappa\approx 3.5$, this yields a
lower bound of approximately 13 TeV on $\Lambda$.

Typically, in addition to the operators in eqn.~(\ref{ops1}) there will
be flavor-changing operators which are products of color-octet
currents\footnote{ Note that it is likely that color must be embedded in
  the flavor interactions in order to avoid possible Goldstone bosons
  \protect\cite{Lane} and large contributions to the 
  $S$ parameter \protect\cite{Spar}.}. 
At least in the vacuum-insertion approximation,
the matrix elements of products of color-octet currents are enhanced
relative to those shown in (\ref{ops1}) by a factor of 4/3 for the LL and
RR operators and a factor of approximately 7 for the LR operator.
Furthermore, because left-handed quarks are weak doublets 
flavor physics associated with the $c$-quark mass may also contribute
to $\Delta S = 2$ interactions. If so, one would replace $m_s$ with
$m_c$ in eqn.~(\ref{eq:fcncbound}), yielding a lower bound on $\Lambda$
of order 20$\sqrt{\kappa}$ TeV.

Furthermore, in the absence of additional superweak
interactions to give rise to CP-violation in $K$-mixing ($\varepsilon$),
the flavor interactions responsible for the $s$-quark Yukawa
couplings must violate CP at some level. In this case the
the bounds on the scale $M/g$ are yet stronger. Recalling that
\beq
 {\rm Re}\, \varepsilon \approx  { {\rm Im M_{12}} \over {2\, \Delta M}} 
 \lae 1.65\,\times\,10^{-3}\, ,
\eeq
and assuming that there are phases of order 1 in the operators shown
in eqn. (\ref{ops1}), we find the bound
\beq
\frac{M}{g}
\gae 3.5 \times 10^3 \, {\rm TeV}~,
\label{fcncmboundcp}
\eeq
yielding a lower bound on $\Lambda$ of order 120$\sqrt{\kappa}$ TeV.
For these reasons, the bounds from 
eqn. (\ref{eq:fcncbound}) may be conservative.

A similar analysis of the link between the $b$-quark mass and
$B_d - \overline{B}_d$ mixing yields the bounds
\beqa
& &  \! \!\left(\frac{M}{g}\right)_{ \!  {\rm LL,RR}} 
\gae 
f_B V_{td} \left( \frac{2  m_B B_B}{3 \,\Delta m_B }\right)^{\! 1/2}
\approx  6.5 \, {\rm TeV} 
\\ [2mm]
& &  \! \!\left(\frac{M}{g}\right)_{ \! {\rm LR}} 
\gae 
f_B V_{td} \left\{ \frac{m_B B_B^{\prime}}{3 \, \Delta m_B } 
\left| \frac{m_B^2}{(m_b + m_d)^2} - \frac{3}{2} \right| \right\}^{\!
    1/2} 
\approx  1.6 \, {\rm TeV} 
\eeqa
on the interactions associated with generating the $b$-quark Yukawa
couplings. Here we have used $f_B \sqrt{B_B} = 0.2$ GeV, $B_B =
B_B^\prime$, $\Delta m_B \approx 3.3 \times 10^{-10}$ MeV, 
$m_b + m_d =
4.5$ GeV, and have assumed that all $b$-$d$ mixing angles are of order
$V_{td} ={\cal O}(10^{-2})$. Applying eqn.~(\ref{eq:yukawa}) in the
case
of the $b$-quark, we find the weaker bound
\beq
\Lambda \gae  1.9 \, {\rm TeV} 
\sqrt{\kappa\left({m_b\over 4.5\, {\rm GeV}}\right)}~.
\label{eq:fcncb}
\eeq
If the flavor physics associated with $t$-quark mass generation
contributes to $\Delta B = 2$ interactions, one should replace $m_b$
with
$m_t$, yielding a bound of order 
12$\sqrt{\kappa}$ TeV. 

Studying the process $b \to s \gamma$ gives no further constraint on the
scale of new physics at present.  The uncertainty in the hadronic matrix
elements is about 25\% \cite{buras}, whereas the direct correction to
the standard model rate for $b \to s \gamma$ from heavy gauge
bosons, such as in ETC models, is about 10 \% \cite{bsg}.

\subsection{Flavor-Changing Neutral Currents: $\Delta C$}

Usually, the strongest constraints on nonstandard physics from
flavor-changing neutral currents come from processes involving Kaons
or $B$-mesons, like those considered above.  In the present case,
however, the constraints from $D^0 - \overline{D}^0$ mixing are also
important because the $c$-quark is heavier than the $s$-quark, while the
$u-c$ mixing is as large as the $d-s$ mixing.  

Again, there are contributions to $D$-meson mixing from the
color-singlet products of currents analogous to those in
eqn.~(\ref{ops1}). The purely left-handed or right-handed
current-current operators yield
\beq
\left(\frac{M}{g}\right)_{ \! {\rm LL,RR}} 
\gae 
f_D\left( \frac{2  m_D B_D}{3 \Delta m_D }\right)^{\! 1/2}
\cos \theta_{L,R}^c \sin \theta_{L,R}^c \approx 120 \, {\rm TeV} ~,
\eeq
where we have used the limit on the neutral $D$-meson mass difference,
$\Delta m_D \lae 1.4 \times 10^{-10}$ MeV \cite{data},
and $f_D \sqrt{B_D} = 0.2$ GeV, $\theta_{L,R}^c \approx \theta_C$.
The bound on the scale of the underlying strongly-interacting
dynamics follows from eqn.~(\ref{eq:yukawa}): 
\beq
\Lambda \gae 11 \, {\rm  TeV}
\sqrt{\kappa\left({m_c\over 1.5\, {\rm GeV}}\right)}~,
\label{eq:Dbound}
\eeq
so that $\Lambda \gae 21$ TeV for $\kappa \approx 3.5$.

The $\Delta C = 2$, LR product of color-singlet currents gives a weaker
bound than eqn.~(\ref{eq:Dbound}) but the LR product of color-octet currents,
\beq
{\cal L}_{eff} = - \,
\cos \theta_L^c \sin \theta_L^c \cos \theta_R^c \sin \theta_R^c
\frac{g^2}{M^2}
( \overline c_L \gamma^{\mu} T^a u_L)
(\overline c_R \gamma_{\mu} T^a u_R) ~,
\label{ops2}
\eeq
where $T^a$ are the generators of $SU(3)_C$, gives a stronger bound:
\beqa
\left(\frac{M}{g}\right)_{ \! {\rm LR}} & \gae &
\frac{4 f_D}{3(m_c + m_u)} \left( \frac{m_D^3 B_D^\prime}{\Delta
  m_D}\right)^{\! 1/2}
(\cos \theta_L^c \sin \theta_L^c \cos \theta_R^c \sin \theta_R^c)^{\!
  1/2}
\\ [2mm]
& \approx & 240 \, {\rm TeV} \left({1.5\, {\rm GeV}\over m_c}\right)~,
\eeqa
corresponding to 
\beq
\Lambda \gae 22 \, {\rm  TeV}
\sqrt{\kappa\left({1.5\, {\rm GeV}\over m_c}\right)}~.
\label{eq:DDbound}
\eeq

\subsection{Corrections to $Z\to b\bar{b}$}

As noted in section 2, the flavor interactions typically produce
non-universal corrections to the couplings of the the $Z$ to ordinary
fermions, eqn.~(\ref{delg}). In conventional models, where $SU(2)_W$ is
not embedded in the new interactions, the effect of these
interactions is to {\it decrease} the width of the $Z$ to each fermion.
Because left-handed quarks transform as weak doublets and the
left-handed $b$-quark is (predominantly) in a doublet with the
$t$-quark, the flavor interactions associated with
the top-quark mass \cite{CSS} can cause potentially important
corrections to the $Zb\bar{b}$ coupling.  

The decay width of the $Z$ into $b$-quarks is most conveniently measured
in terms
of the ratio, $R_b$, of the $b$-quark partial width to the hadronic
partial width.  
A change $\delta g^b_L$ of the $Z$ boson's coupling to $b$-quarks would result
in a change in $R_b$ relative to the standard model of 
\beq
\delta R_b \approx R_b(1- R_b) \frac{2 g_L^b \delta g_L^b}
{(g_{L}^b)^2 + (g_{R}^b)^2} \approx - 0.774 \, \delta g_L^b~\ \ .
\label{rb}
\eeq
From eqns.~(\ref{eq:yukawa}), (\ref{delg}) and (\ref{rb}) we find
\beqa
\Lambda & = & \left( \frac{m_t v}{2\sqrt{2}} \, 
\frac{0.774}{(-\delta R_b)}\kappa
\right)^{\! 1/2} 
\\ [2mm]
& \approx & 0.11 \, 
{\rm TeV} \left( \frac{\kappa}{-\delta R_b}\right)^{\! 1/2}~.
\eeqa

Note that a conventional model only accommodates a {\it decrease} in
$R_b$. For this reason the limits we can place on $\Lambda$ are
extremely sensitive to the bounds on negative values of $\delta R_b$.
The current ``best fit'' to LEP and SLD data yields the value
\cite{lepewwg}
\begin{displaymath}
R_b = 0.2178\, \pm\, 0.0011~,
\end{displaymath}
as compared to the standard model prediction (for $m_t=
172\, \pm 6$ GeV \cite{cdf}) of
\begin{displaymath}
R^{\rm sm}_b = 0.2158~.  
\end{displaymath}
These imply 
\beq
\delta R_b = 0.0020 \pm 0.0011~.
\eeq
At 95\% confidence level $-\delta R_b|_{2\sigma} \leq 0.0002$,
corresponding to
\beq
\Lambda \gae 7.7 \, {\rm TeV} \sqrt \kappa ~,
\label{eq:zbbound}
\eeq
(14 TeV for $\kappa\approx 3.5$)
whereas at 99.7\% confidence level 
$-\delta R_b|_{3\sigma} \leq  0.0013$, corresponding to 
\beq
\Lambda \gae 3.1 \, {\rm TeV} \sqrt \kappa~.
\eeq
Given the sensitivity to the confidence
level, we view the bound in eqn.~(\ref{eq:zbbound}) as less
``robust'' than the bounds from $K-\overline{K}$ or
$D-\overline{D}$ mixing
(eqns.~(\ref{eq:fcncbound}) and (\ref{eq:Dbound})).

\section{Higgs Mass Limits}
\setcounter{equation}{0}

Because of triviality, a lower bound on the scale $\Lambda$ yields an
upper limit on the Higgs boson's mass. A rigorous determination of this
limit would require a nonperturbative calculation of the Higgs mass in
an $O(4)$-symmetric theory subject to the constraint on $\Lambda$.  Here
we provide an estimate of this upper limit by naive extrapolation of the
lowest-order perturbative result\footnote{The naive perturbative bound
  has been remarkably close to the non-perturbative estimates derived
  from lattice Monte Carlo calculations \cite{refbound}.}.  
Integrating the lowest-order beta function for
the Higgs self-coupling $\lambda$,
\beq 
\beta(\lambda) = \mu{d\lambda\over d\mu} = {3\over
  2\pi^2}\lambda^2 +\ldots~, 
\eeq 
we find 
\beq 
{1\over \lambda(\mu)} -
{1\over \lambda(\Lambda)} = {3\over 2\pi^2} \ln{\Lambda\over \mu} ~.
\eeq 
Using the relation $m^2_H = 2\lambda(m_H) v^2$ we find the relation
\beq 
m^2_H \ln\left({\Lambda\over m_H}\right)\le {4\pi^2 v^2 \over 3}~.
\label{thisw}
\eeq 

The lower bounds on $\Lambda$ from section 3 may be combined with
eqn.~(\ref{thisw}) to yield corresponding upper bounds on $m_H$.  The
bound $\Lambda > 13$ TeV given by the contribution of the $\Delta S = 2$
product of color-singlet currents to the $K_L - K_S$ mass difference,
eqn.~(\ref{eq:fcncbound}), in the case $\kappa \approx 3.5$, results in
the limit\footnote{If $\kappa \approx 4\pi$, $\Lambda$ would have to be
  greater than 24 TeV, yielding an upper limit on the Higgs boson's mass
  of 450 GeV. If $\kappa \approx 1$, $\Lambda$ would be greater than 6.8
  TeV, yielding the upper limit $m_H \laef 570$ GeV.} $m_H \lae 490$
GeV.  The bound $\Lambda \gae 21$ TeV, given by the contribution of the
$\Delta C = 2 \,$, LL or RR product of color-singlet currents to the
neutral $D$-meson mass difference, eqn.~(\ref{eq:Dbound}), yields $m_H
\lae 460$ GeV.  Limits from the contributions of color-octet currents or
from the relationship between $m_c$ and $\Delta m_K$ would be even more
stringent.  Finally, if the flavor interactions responsible for the
$s$-quark Yukawa coupling also generate CP-violation in Kaon mixing and
there are phases of order 1 in the interactions in eqn. (\ref{ops1}),
the resulting bound $\Lambda > 220\ $TeV would yield a Higgs mass limit
of 350 GeV.

\section{Conclusions}

Because of triviality, theories with a heavy Higgs boson are effective
low-energy theories valid below some cut-off scale $\Lambda$.  The
underlying high-energy theory must include flavor dynamics at a scale of
order $\Lambda$ or greater in order to produce the different Yukawa
couplings of the Higgs to ordinary fermions.  This flavor dynamics will
generically give rise to flavor-changing neutral currents and
non-universal corrections to the decay $Z\to b\bar{b}$. In this note we
showed that satisfying the experimental constraints on extra
contributions to $\Delta m_K$, $\Delta m_D$, and $R_b$  requires that the
scale of the associated flavor dynamics exceed certain lower bounds. At
the same time, the new physics must provide sufficiently large Yukawa
couplings to give the quarks their observed masses.  In order to give
rise to a sufficiently large $s$-quark Yukawa coupling, we showed that
$\Lambda$ must be greater than of order 13 TeV, while in the case of the
$c$-quark the bound is even more stringent, $\Lambda \gae 21$ TeV.  For
theories defined about the infrared-stable Gaussian fixed-point, we
estimated that this lower bound on $\Lambda$ yields an upper limit of
approximately 460 GeV on the Higgs boson's mass, independent of the
regulator chosen to define the theory.


\vspace{12pt} \centerline{\bf Acknowledgments} \vspace{10pt}

The authors thank H. Georgi for useful comments and for emphasizing the
potential relevance of CP violation.
R.S.C. and E.H.S. thank Koichi Yamawaki and the organizers of the 1996
International Workshop on Perspectives of Strong Coupling Gauge Theories
(SCGT 96) held in Nagoya, Japan from 13-16 November, 1996 for holding a
stimulating conference where this work was begun. E.H.S. acknowledges
the support of the NSF Faculty Early Career Development (CAREER)
program, the DOE Outstanding Junior Investigator program and the JSPS
Invitation Fellowship program. {\em This work was
  supported in part by the National Science Foundation under grants
  PHY-9057173 and PHY-9501249, and by the Department of Energy under
  grant DE-FG02-91ER40676.}


\end{document}